\documentclass[conference]{IEEEtran}
\IEEEoverridecommandlockouts
\usepackage{cite}
\usepackage{graphicx}
\usepackage{booktabs}
\usepackage{array}
\usepackage{url}
\usepackage{amsmath}
\usepackage{amssymb}
\usepackage{pifont}
\usepackage{threeparttable}
\usepackage{multirow}
\usepackage{makecell}
\usepackage[table]{xcolor}
\newcommand{\cmark}{\textcolor{green}{\ding{51}~}}%
\newcommand{\xmark}{\textcolor{red}{\ding{55}~}}%

\begin{document}

\title{SoK: Cross-Chain Transaction\\ Identification and Matching}

\author{
\IEEEauthorblockN{Hang Zheng\IEEEauthorrefmark{1},
Qishuang Fu\IEEEauthorrefmark{1}\thanks{Corresponding author: Qishuang Fu
(qishuang.fu@monash.edu).},
Joseph Liu\IEEEauthorrefmark{1},
Qin Wang\IEEEauthorrefmark{2},
Weiqing Wang\IEEEauthorrefmark{1},
and Tsz Hon Yuen\IEEEauthorrefmark{1}}
\smallskip
\IEEEauthorblockA{\IEEEauthorrefmark{1}\textit{Monash University, Australia}
}
\IEEEauthorblockA{\IEEEauthorrefmark{2}\textit{CSIRO, Australia}
}
}

\maketitle

\begin{abstract}
Cross-chain bridges, instant cryptocurrency exchanges, and centralized
cross-ledger platforms move assets across an increasingly multi-chain ecosystem.
However, these systems have repeatedly become targets of high-value attacks and channels for cross-chain money laundering.
Cross-chain transactions are substantially harder to analyze than
single-chain transactions: 
no single ledger records an entire cross-chain transfer,
its evidence is scattered across the source chain, the destination chain, and
off-chain systems, and the availability and reliability of that evidence vary widely across systems.
In this paper, we present a systematization of knowledge (SoK) on cross-chain transaction identification and matching. First, we classify deposit and withdrawal identification methods into four approaches
and transaction matching methods into
three mechanisms: deterministic identifier matching, field-constraint
heuristics, and model-assisted matching. We find that their applicability and reported
performance are shaped mainly by the evidence the underlying system exposes,
and we further examine how matched pairs support downstream attack detection and fund
tracing. Second, we assess the availability of existing datasets and
artifacts, finding that fewer than half remain obtainable, and
distill three artifact failure modes. Finally, we outline four open challenges
toward auditable, reproducible, and actionable cross-chain analysis.
\end{abstract}
\smallskip

\section{Introduction}
\label{sec:intro}

Over the past few years, the blockchain ecosystem has evolved from isolated
chains into a multi-ledger environment comprising Layer-1 networks, Layer-2 networks, and sidechains, and assets
increasingly move across chains through bridges and cross-ledger exchange
services \cite{gudgeon2020sok,qi2025sok,zhang2024security,li2025blockchain}. 
As of July 2026, the total cryptocurrency market capitalization is
 \$2.2 trillion~\cite{coingecko2026marketcap}, and monthly cross-chain
transfer volume exceeds \$12 billion~\cite{defillama2026bridges}. 
Cross-chain services have also repeatedly become targets of major attacks. Cumulative losses from bridge attacks have reached approximately \$2.8 billion, accounting for about 40\% of the value stolen in Web3~\cite{defillama2026hacks,chainalysis2022bridges,wang2022exploring}. 
Cross-chain services are also used to obscure fund flows. Elliptic~\cite{elliptic2025crime}
estimates that more than \$21.8 billion of illicit and high-risk funds have moved
through DEXs, bridges, and swap services without know-your-customer (KYC) checks.
The Lazarus Group, for example, laundered
the Bybit proceeds by rapidly hopping across chains~\cite{trm2026crime}. Consequently, cross-chain transactions have become an important object of study for understanding fund flows and
the attack surface of the multi-chain ecosystem.

However, cross-chain transactions are substantially harder to analyze than
single-chain transactions, because no single blockchain records an entire cross-chain transfer. 
The evidence is scattered across the source chain, the destination
chain, and off-chain systems, and its availability and reliability vary considerably across platforms. Some bridges record an identifier that uniquely links the two ends of a transfer. In other systems, the deposit and withdrawal appear on-chain as two ordinary transactions sharing only time, amount, and addresses.

Existing studies have developed a range of methods for cross-chain transaction matching.
These methods first identify cross-chain-related transactions among
massive on-chain data, and then associate the two ends into transaction pairs
or paths. The resulting matches provide important inputs to downstream applications, including attack detection and fund tracing.
Although cross-chain transaction matching has attracted increasing research attention, existing SoKs on cross-chain
systems~\cite{zamyatin2021sok,zhang2024cross,augusto2024sok,wang2023exploring,notland2026sok,lee2023bridgehacks,belenkov2025bridgehacks}
focus on protocol design, attack surfaces, and architectural flaws, and they do not
systematize how transaction-level evidence is used to match cross-chain
transactions. The closest effort, the survey of transaction tracing
techniques \cite{kumar2025tracingsurvey}, centers on
single-chain tracing and does not compare the evidence strength of matching
mechanisms.

To fill this gap, we systematize research on cross-chain transaction
identification and matching. 

Our contributions are as follows.
\begin{itemize}
\item We organize deposit and withdrawal identification into four approaches
  and transaction matching into three mechanisms (deterministic identifier
  matching, field-constraint heuristics, and model-assisted matching),
  comparing their evidence sources, applicability conditions, and failure
  modes (\S\ref{sec:identification} and~\S\ref{sec:matching}).
\item We compare public datasets and artifacts and verify their actual
  availability one by one (\S\ref{sec:datasets}).
\item We distill four open challenges: interpreting missing counterparts under
  partial evidence; one-to-many and many-to-many matching; ground truth,
  benchmarks, and artifact availability; and verifiable use of models and
  agents (\S\ref{sec:challenges}).
\end{itemize}

\textit{Scope and paper selection:}
This SoK targets transaction data produced by deployed cross-chain systems,
covering contract bridges (including canonical L1--L2 bridges and cross-chain
messaging protocols) as well as instant cryptocurrency exchanges and other
centralized cross-ledger platforms. 

We include works published between
2019 and 2026 that analyze transaction data from such systems, collected through
keyword search over major security, measurement, and blockchain venues and
citation snowballing from seed papers. Works on protocol design, contract-level
vulnerability analysis (static analysis and fuzzing), and single-chain-only
transaction analysis are excluded from the survey and cited only where
relevant. Detection-only works are not part of the core taxonomy and are cited
only where they show downstream use or evidence limits.

\section{Background}
\label{sec:background}

\emph{Cross-chain} refers to the techniques that reliably move assets, data, or
messages between different blockchain networks. Chains such as
Bitcoin~\cite{nakamoto2008bitcoin}, Ethereum~\cite{wood2014ethereum}, and BNB
Chain~\cite{bnbchain2026docs} are independent ledgers that cannot interoperate
directly. Cross-chain techniques address this problem, and the services that
implement them are \emph{cross-chain systems}.

\subsection{Cross-Chain Systems and Their Evidence}
\label{sec:bg-systems}

This paper groups cross-chain systems by the form of service they provide,
because the forms differ in what they leave on chain, and hence in what evidence
is available for analysis. The first form, the smart-contract bridge, runs a
contract on each of the two chains and moves assets through this pair, so both
ends leave contract events. Bridges differ in who confirms that the source-chain
deposit really happened~\cite{belchior2022survey}: a notary or committee, an
optimistic challenge window, or the destination chain itself via a light client
or proof.
The second form covers instant cryptocurrency exchanges, ShapeShift-style
cross-ledger swap services~\cite{hu2024instant}, centralized exchanges, and other
virtual asset service providers: they all run the exchange itself off chain, so
the public chains see only the deposit and withdrawal transfers themselves.
The third form, the
\emph{aggregator}, runs no exchange of its own: it hands a transfer to whichever
bridges or swap services currently offer the best rate, sometimes splitting it
among several, so one deposit no longer needs to correspond to one withdrawal.

\begin{figure*}[t]
\centering
\includegraphics[width=\textwidth]{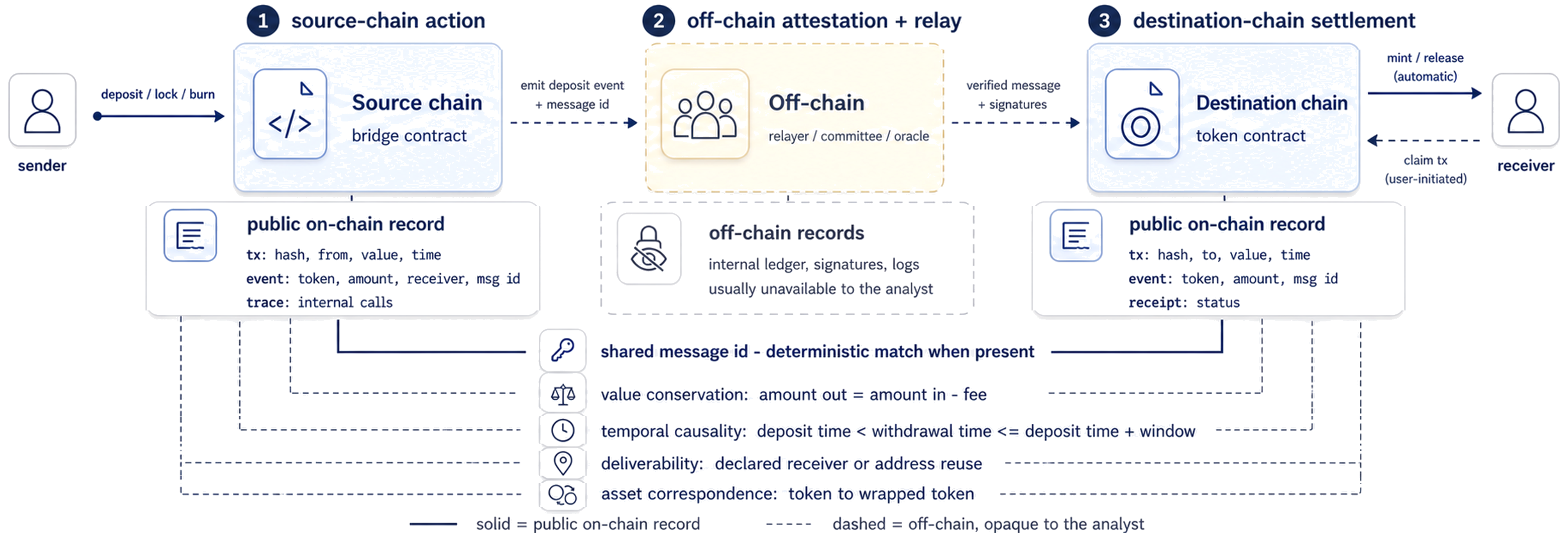}
\caption{Lifecycle of a cross-chain transfer and where its evidence is recorded.
A transfer moves through a source-chain action, an off-chain attestation and
relay, and a destination-chain settlement. Each stage leaves a different kind of
evidence, and depending on the system, the two ends may be linked by explicit
identifiers, platform records, or inferred business invariants.}
\label{fig:lifecycle}
\end{figure*}

For a contract-based bridge, one user-level cross-chain transfer can be abstracted
into three stages (Fig.~\ref{fig:lifecycle}): the user initiates the transfer on
the source chain by locking, burning, or depositing the asset; an off-chain
component, typically the bridge's verifier or a relayer, attests to that event
and passes a message; and the destination-chain
contract settles the transfer by minting, releasing from custody or a liquidity
pool, or waiting for the user to claim. A bridge's \emph{settlement
mechanism}~\cite{han2023crosschain,zhang2026atomxross,geng2023subsidy} takes one of three forms: lock the asset and
mint a wrapped copy on the other chain (lock-and-mint), burn it on the source
chain and mint it on the destination (burn-and-mint), or pay out of pre-funded
pools on both chains (liquidity pools). The two directions of the same bridge can also differ: some
withdrawals need a second, user-timed claim transaction on the original chain.

An instant exchange or cross-ledger service follows the same three stages, but the
attestation and the settlement are hidden inside the platform: the
user sends an ordinary transfer to a deposit address that the platform assigns
off-chain; the platform matches the request internally; and on the other chain it
pays the user from its funds. Because no bridge contract is involved, neither
transaction writes any bridge event, and the two look like unrelated ordinary
transfers. 
The link is still there, however: a swap executes one
quoted request at a time, so each deposit is usually paid out by one withdrawal
within minutes. 
Any difference between the two amounts comes only from the quoted
rate and the fee. A centralized exchange
breaks this coupling too: a deposit only tops up an account balance, the user may trade, wait,
or split the money, and withdrawals leave from the same shared hot wallets, so a
traced fund path often stops at the exchange's deposit address.

Cross-chain analysis relies first on on-chain records: the transaction itself, the
event logs written by bridge and token contracts, the execution traces that show
what actually happened, and the contract ABI and source code that give fields
their business semantics. How much these records reveal varies by bridge: some bridges write an identifier
usable directly for pairing into their events, while others expose only generic
fields such as receiver, token, amount, and timestamp.

Beyond on-chain records, existing work also uses off-chain and external evidence. The first
is the platform's records: bridge ledgers, official bridge explorers
(transaction-lookup websites analogous to a block explorer such as Etherscan), and
platform APIs. Where a platform exposes them, these records name the transactions on
both ends of a transfer together with its status, amount, and time. The second is
external labels: entity tags that name an address's owner (an exchange, a service, a
known attacker), incident reports, and third-party risk annotations. Analyses also
consult market prices and exchange rates when amounts must be compared across
assets.

\subsection{Analysis Tasks}
\label{sec:bg-tasks}

Cross-chain transaction analysis involves two core tasks.
The first is \emph{deposit and withdrawal identification}: out of the full
transaction stream of a chain, decide which transactions are cross-chain deposits
or withdrawals at all. 

The second is \emph{transaction matching}: given the
deposits on one chain and the withdrawals on another, decide which two belong to
the same cross-chain operation. A matched deposit--withdrawal pair is called a
cross-chain transaction (cctx)~\cite{augusto2025xchainwatcher,augusto2024sok}.
\emph{Tracing}, which strings matched pairs and ordinary transfers into an
end-to-end fund path, and \emph{anomaly detection}, which flags the transactions
or accounts that break a system's rules or deviate from normal behavior, build on
these results. This SoK discusses both only as downstream applications of
recovered cross-chain evidence.

\section{Identifying Cross-Chain Transactions}
\label{sec:identification}

Cross-chain analysis starts from identification: before any pair can be matched,
the deposits and withdrawals must first be found among a chain's transactions.
Existing work identifies cross-chain deposits and withdrawals in four ways:
official-record lookup, signature matching, address aggregation, and learned
classification.

\emph{Official-record lookup.} The ledger of a bridge or platform already stores the
hash and status of deposit and withdrawal transactions. Yan et
al.~\cite{yan2025empirical} obtain bridge ledgers through the transaction-info
interfaces of four bridges, Yousaf et al.~\cite{yousaf2019tracing} poll
ShapeShift's public interface, and CLTracer~\cite{zhang2022cltracer} calls its
transaction-status interface (\texttt{txstat}) to verify candidate deposits. This
class depends entirely on whether the platform offers an interface: among the
thirty bridges surveyed by Yan et al., only four provide a usable API.

\emph{Signature matching.} An analyst audits the bridge contract's source code,
enumerates the dedicated events and function selectors of its deposit and
withdrawal functions, and scans on-chain logs. The Ethereum--Polygon
tracing~\cite{yan2025tracing} identifies deposits through
\texttt{LockedEther}-family events, Jigsaw~\cite{hu2024jigsaw} compiles such
signature lists for thirteen bridges, and CONNECTOR~\cite{lin2025connector},
XChainWatcher~\cite{augusto2025xchainwatcher}, and XScope~\cite{zhang2022xscope}
maintain equivalent decoders. This class requires the contract to be open source,
and a bridge upgrade that renames an event invalidates the hand-written rules.

\emph{Address aggregation.} A centralized service leaves
only ordinary transfers on chain, but its fund management forms a funnel-shaped
topology in which many one-off deposit addresses aggregate into a few central
addresses or hot wallets. CLTracer locates the shared central address from seed
transactions and collects in reverse the transactions paid into it, recovering
1.77 million deposits; the ICE study~\cite{hu2024instant} confirms a service's hot
wallet with real exchange requests and infers deposit addresses backward from the
aggregation transactions. This class presupposes a central-aggregation
architecture, and CLTracer's survey of nineteen alternative platforms maps its
boundary: roughly half can be traced the same way, while the rest either hand the
received funds to traditional exchanges (the aggregation edge leads into an
exchange wallet) or, like the privacy-conscious Flyp.me, leave deposits unspent
(the edge never appears).

\emph{Learned classification.} This class casts identification as a classification
task over a transaction's
execution structure and textual semantics. XSema~\cite{zheng2024xsema} classifies
transactions into deposit, withdrawal, and non-cross-chain from the transaction's
transfer structure and event-log text.
CONNECTOR combines textual and structural features: it embeds the function text so
that deposit functions named differently across bridges read as the same intent,
and builds a call graph of the internal calls and asset transfers the deposit
triggers. The two feature families together identify nearly 100\% of deposit
behaviors, while structure alone reaches 74.43\%. Both works build their training
labels from official bridge explorers, so a learning-based method still depends on
platform cooperation. The dependence merely moves from inference time to training
time. The real value of this class lies in cross-bridge generalization: trained on
four of ten bridges,
XSema reaches a 94.85\% macro F1 on the six unseen during training.

\section{Matching Cross-Chain Transactions}
\label{sec:matching}

Matching takes over from identification: it decides which deposit and which
withdrawal complete the same transfer across chains. Existing methods fall into
three classes by the evidence they rely on: deterministic identifier matching,
field-constraint heuristics, and model-assisted matching
(\S\ref{sec:deterministic}--\S\ref{sec:model-assisted}).
Table~\ref{tab:matching} summarizes the works.


\providecommand{\graycmidrule}[1]{%
  \arrayrulecolor{black!15}%
  \cmidrule[0.3pt](lr){#1}%
  \arrayrulecolor{black}%
}

\begin{table*}[t]
\centering
\caption{Cross-Chain Deposit/Withdrawal Identification and Matching Works}
\label{tab:matching}
\scriptsize
\setlength{\tabcolsep}{4pt}
\renewcommand{\arraystretch}{1.1}
\begin{tabular}{@{}c l p{0.6cm}  ccl l p{0.6cm} p{0.6cm}  p{0.6cm}  c p{0.6cm} @{}}
\toprule
 &  & & & & \multicolumn{4}{c}{Matching} & \multicolumn{3}{c}{Outcome} \\
\cmidrule(lr){6-9} \cmidrule(lr){10-12}
 & Method & Yr & Sys & Identifi. & Mech & Pairing evidence & Addr & Gen. & Pairs & Performance & Code \\
\midrule

\multirow{1}{*}{\textit{Identification only}} & 
XSema~\cite{zheng2024xsema} & 2024 & BR & L & -- & -- & -- & \cmark & -- & 94.85\% ident.\ F1 & \textcolor{teal}{$\circ$} \\
\midrule

\multirow{8}{*}{\makecell{\textit{Deterministic} \\ \textit{identifier matching}}} 
& CLTracer~\cite{zhang2022cltracer} & 2022 & ICE & R,A & D$_\text{off}$ (H) & platform record & -- & \xmark & -- & -- & \textcolor{teal}{$\circ$} \\
\graycmidrule{2-12}
& Empirical Study~\cite{yan2025empirical} & 2025 & BR & R & D$_\text{off}$ & platform record & -- & \xmark & 543K & -- & \textcolor{orange}{$\bullet$} \\
\graycmidrule{2-12}
& Jigsaw~\cite{hu2024jigsaw} & 2024 & BR & E & D$_\text{on}$ (H) & on-chain identifier & -- & \xmark & -- & 99.8\% match rate & \textcolor{teal}{$\circ$} \\
\graycmidrule{2-12}
& XChainWatcher~\cite{augusto2025xchainwatcher} & 2025 & BR & E & D$_\text{on}$ & on-chain identifier & -- & \xmark & 81K & -- & \textcolor{orange}{$\bullet$} \\
\graycmidrule{2-12}
& Arbitrage~\cite{oz2025arbitrage} & 2025 & L2 & E & D$_\text{on}$ (H) & on-chain identifier & reuse & \xmark & -- & -- & \textcolor{teal}{$\circ$} \\
\graycmidrule{2-12}
& XChainDataGen~\cite{augusto2025xchaindatagen} & 2025 & BR & E & D$_\text{on}$ & on-chain identifier & -- & \xmark & 11.29M & -- & \textcolor{orange}{$\bullet$} \\
\midrule

\multirow{5}{*}{\makecell{\textit{Field-constraint} \\ \textit{heuristic matching}}} 
& Yousaf et al.~\cite{yousaf2019tracing} & 2019 & ICE & R & H (D$_\text{off}$) & amt/time & reuse & \xmark & 1.38M & -- & \textcolor{teal}{$\circ$} \\
\graycmidrule{2-12}
& ICE Study~\cite{hu2024instant} & 2024 & ICE & A & H & addr/amt/time & reuse & \xmark & 194K & 80.8--91.9\% match rate & \textcolor{teal}{$\circ$} \\
\graycmidrule{2-12}
& Eth-Polygon~\cite{yan2025tracing} & 2025 & L2 & E & H & addr/amt/time/token & reuse & \xmark & 2M & 67.55--92.78\% match rate & \textcolor{teal}{$\circ$} \\
\graycmidrule{2-12}
& CONNECTOR~\cite{lin2025connector} & 2025 & BR & E,L & H & time/token/fee & decl. & \xmark & 24K & 95.95\% match rate & \textcolor{orange}{$\bullet$} \\
\graycmidrule{2-12}
& LOCARD~\cite{yu2026locard} & 2026 & BR & R,E & H (Agent) & amt/time/fee & -- & \xmark & 151K & -- & \textcolor{orange}{$\bullet$} \\
\midrule

\multirow{2}{*}{\makecell{\textit{Model-assisted} \\ \textit{matching}}} 
& ABCTRACER~\cite{lin2025abctracer} & 2025 & BR & L & M & learned + explicit clues & decl. & \cmark & 29K & 91.75\% F1 & \textcolor{orange}{$\bullet$} \\
\graycmidrule{2-12}
& ConneX~\cite{liang2025connex} & 2025 & BR & E,L & M (H) & LLM-selected fields + rules & decl. & \cmark & -- & 97.46\% F1 & \textcolor{teal}{$\circ$} \\
\bottomrule
\end{tabular}

\smallskip
\begin{minipage}{\textwidth}
\scriptsize
 Sys: BR = contract bridge, ICE = instant exchange, L2 = L1--L2
bridge. Identification: R = record lookup, E = signature matching, A = address
aggregation, L = learned classification.
Mech: D$_\text{off}$/D$_\text{on}$ = deterministic off-/on-chain, H = field
heuristic, M = model-assisted; parentheses mark a secondary mechanism. Field
lists are ordered addr, amt, time, token, fee.
Addr: decl. = protocol-declared receiver, reuse = same-address assumption.
Gen.: \cmark = demonstrated reduced per-bridge adaptation
(unseen-bridge tests or automatic field discovery), \xmark =
per-bridge manual rules.
Pairs = cross-chain transaction pairs the work reports
collecting or matching; -- = not reported at pair granularity.
Performance: match rate = share of transactions successfully matched,
with the denominator and correctness check defined by each work;
F1 = precision--recall harmonic mean on labeled pairs;
ident.\ = identification task. Figures are as stated by each work and
not comparable across papers.
Code: \textcolor{orange}{$\bullet$} = code released, \textcolor{teal}{$\circ$} = not released; dataset
availability is in Table~\ref{tab:datasets}.
\end{minipage}
\end{table*}

\subsection{Deterministic Identifier Matching}
\label{sec:deterministic}

Deterministic identifier matching relies on an association clue that the system or
platform records explicitly. By where the identifier resides, existing methods split
into two subclasses: off-chain platform records, which give the pairing directly,
and on-chain event identifiers, which the bridge contract writes into events as a
message identifier, a source transaction hash, or a deposit identifier.

Off-chain platform records provide already-matched transactions: a ledger entry or
an API response (\S\ref{sec:identification}) names the transactions on both
ends. The remaining work is to pull the named transactions from the two chains and
to check that the record and the chain agree. Such records fail in three ways. The
first is a correctness failure: when Yan et al.~\cite{yan2025empirical} compared
four bridges' ledgers with the chains, 9,956 transactions differed in amount and 308
were marked successful but had failed on chain. The second is an availability
failure: after
Multichain collapsed in 2023, the works that depend on its official explorer data,
CrossAAD~\cite{crossaad2024} and GMMCCT~\cite{gmmcct2024}, can no longer be
reproduced. The third is a coverage failure: a
platform record covers only activity inside that platform, and
CLTracer~\cite{zhang2022cltracer} found that phishing- and hacker-related addresses
completed their last cross-ledger transaction the day before ShapeShift's mandatory
KYC took effect. ShapeShift's usage then declined drastically. Off-chain records
give a strong pairing clue within a platform, but their coverage decays as platform
policy and user behavior change.

On-chain event identifiers are unique fields a bridge contract writes into its
event logs for the correctness, so that the same message cannot be executed
twice. Because they live on chain, anyone can re-verify them at any time,
and they do not disappear when a platform does. Jigsaw~\cite{hu2024jigsaw} pairs
with such fields on ten of thirteen bridges (the \texttt{srcTxHash} in Multichain's
\texttt{LogAnySwapIn}, for example) and, with a heuristic fallback for the other
three, reaches a 99.8\% overall matching rate on about eighty million transactions,
the largest matching study we survey.
XChainWatcher~\cite{augusto2025xchainwatcher} and the cross-chain arbitrage
study~\cite{oz2025arbitrage} join on the same kind of key (a \texttt{deposit\_id},
a message number), and XChainDataGen~\cite{augusto2025xchaindatagen} extracts
11.29M cctx from five protocols across eleven chains using such identifiers
alone, released as a public dataset (\S\ref{sec:datasets}).

\smallskip
\emph{\underline{Insight 1}: Beyond strength, evidence differs in verifiability and durability;
the most dependable pairing evidence is a publicly verifiable join key on chain.
Both deterministic sources give the pairing outright, but a record can be wrong, go
offline, or lose coverage, whereas a key, wherever the contract writes it, stays on
chain for anyone to re-check.}

\subsection{Field-Constraint Heuristics}
\label{sec:heuristics}

When neither record nor identifier exists, only the ordinary fields of a transfer
remain (addresses, amounts, timestamps, tokens), and field-constraint heuristics
match with these fields. Such heuristics first narrow the candidate set by address:
the receiver recorded in the deposit event, the same user address appearing on both
chains, or an address that platform records associate with the user. When no
address clue is available, they treat every transfer within a time window as a
candidate. They then retain only the candidates whose timing, post-fee amount,
token, and destination chain are consistent with the deposit. The two ends never
agree exactly. The withdrawal arrives later than the deposit and, after fees, for
a smaller amount. Every rule therefore needs a tolerance, and the analyst must
decide its size, such as the width of the time window and the allowed deviation in amount.
Table~\ref{tab:fields} summarizes the evidence constraints matching relies on,
what each checks, and when each breaks.

\providecommand{\grayrule}{\arrayrulecolor{black!15}\specialrule{0.3pt}{1pt}{1pt}\arrayrulecolor{black}}
\begin{table*}[t]
\centering
\caption{Evidence Constraints in Cross-Chain Matching}
\label{tab:fields}
\footnotesize
\setlength{\tabcolsep}{6pt}
\renewcommand{\arraystretch}{1.15}
\begin{tabular}{@{}p{2.6cm} l l p{4cm}@{}}
\toprule
\multicolumn{1}{c}{Evidence} & \multicolumn{1}{c}{What it checks} & \multicolumn{1}{c}{Typical signal} & \multicolumn{1}{c}{Breaks when} \\
\midrule
Protocol identifier & direct pairing & srcTxHash; deposit\_id; message number & bridge writes no key \\
\grayrule
Declared address & named recipient matches & receiver field & field not emitted \\
\grayrule
Reused address & same user on both chains & same EOA or platform address & recipient differs from sender \\
\grayrule
Amount & value conservation & exact amount; post-fee amount; FX-converted amount & units, fees, and rates vary \\
\grayrule
Time & deposit precedes withdrawal & time window; finality bound & user-timed claims stretch window \\
\grayrule
Asset and route & token and chain correspondence & token mapping; token ID & hand-kept mappings go stale \\
\bottomrule
\end{tabular}

\smallskip
\begin{minipage}{\textwidth}
\footnotesize
Representative settings: amount tolerance
0.5\%~\cite{oz2025arbitrage}, fee $\in[0,3\%]$~\cite{lin2025connector},
20\%~\cite{liang2025connex}, FX rate~\cite{yu2026locard}; time window
5\,min~\cite{hu2024instant}, 30\,min~\cite{lin2025connector},
2\,h~\cite{liang2025connex}, 24.2\,min / 6.4\,d~\cite{yan2025tracing}.
Identifier keys: srcTxHash~\cite{hu2024jigsaw},
deposit\_id~\cite{augusto2025xchainwatcher}, message
number~\cite{oz2025arbitrage}. Field location can itself be learned:
NER~\cite{lin2025abctracer}, LLM selection~\cite{liang2025connex}.
\end{minipage}
\end{table*}

The same heuristics perform very differently as the evidence changes. In the
Ethereum--Polygon tracing of Yan et al.~\cite{yan2025tracing}, deposit matching
exceeds 93\%, but the ERC20 withdrawal matching rate is only
67.55\%, because the bridge produces no dedicated event for an ERC20
withdrawal (only a standard \texttt{Transfer}), which makes a bridge withdrawal
indistinguishable at the log level from an exchange withdrawal or an ordinary
transfer; an ERC721 withdrawal, which has a dedicated event and a unique token
identifier, reaches 92.78\%. A withdrawal on this bridge also completes in two
steps: the user burns the wrapped copy on Polygon, then claims the funds on
Ethereum with a second transaction, whenever they choose. The best time window
therefore widens from about 24.2 minutes for deposits to about 6.4 days, leaving
the time constraint with almost no filtering power.

Heuristics also serve as every system's fallback layer: the basic version of Yousaf
et al.~\cite{yousaf2019tracing} works without the platform interface, and
Jigsaw~\cite{hu2024jigsaw} drops to field heuristics on the three bridges that write
no identifier. Among the works that match with fields alone, the ICE
study~\cite{hu2024instant} reaches 80.8\%--91.9\% and
CONNECTOR~\cite{lin2025connector} 95.95\%.
LOCARD~\cite{yu2026locard} runs the same kind of heuristics inside an agent
workflow. Despite
these rates, such heuristics share a weak point: filtering often leaves more than
one candidate standing. Existing works disambiguate with rules chosen by hand, and
each chooses differently: ConneX~\cite{liang2025connex} takes the candidate with
the earliest timestamp, CONNECTOR tightens the time window,
ABCTRACER~\cite{lin2025abctracer} and LOCARD score the candidates, and Yan et al.
keep a match only when exactly one candidate passes.

\subsection{Model-Assisted Matching}
\label{sec:model-assisted}

Model-assisted matching uses learning models in two ways: some works let a
model find the useful fields and clues, and others train a model on data to pair
the two ends directly. Both aim to lower the cost of adapting to bridges where an
explicit identifier is missing and field semantics are heterogeneous.

Model-assisted matching faces two kinds of field heterogeneity. The first is
naming: the useful fields (the destination chain, the receiver address) appear in
every bridge's events, but each bridge names them differently.
ABCTRACER~\cite{lin2025abctracer}, which targets automated, bidirectional tracing
on DeFi bridges, uses named-entity recognition to identify a field by its meaning
(its explicit clues); ConneX~\cite{liang2025connex} is more conservative with its
LLM, which only identifies, among many candidate fields, the five-tuple that may
carry cross-chain semantics, while a deterministic examiner with six rules
completes verification and pairing, for an overall F1 of 0.9746, against only
0.71 when a baseline lets the LLM choose the paired transaction directly. The
second is the clues themselves: what links the two ends differs from bridge to
bridge. On Allbridge, for example, a deposit and its withdrawal carry the same
recipient, nonce, and messenger values. ABCTRACER trains a Siamese network, a model that compares two
transactions as a pair, on known matched pairs to uncover such patterns (its
implicit clues). Combining the two kinds of clue, it reaches a 91.75\%
bidirectional F1 on twelve bridges; trained on only 25\% of the bridges, it still
reaches a 93.95\% F1 on unseen ones. The main value of a learning-based method
thus lies in reducing the manual cost per bridge.

The evaluation of a learning-based method should therefore consider, besides
single-point F1, cross-bridge generalization, manual annotation cost, and whether
the final pairing remains verifiable: F1 alone ranks ABCTRACER's 91.75\% below
Jigsaw's 99.8\% deterministic matching. ConneX's division of labor, in which the
model assists and deterministic rules decide, also holds for agents, programs
that use a large language model (LLM) as their decision core, plan steps,
and call external tools: in LOCARD~\cite{yu2026locard}, agents plan the queries,
collect the evidence, and track the investigation state, and the pairing decision
still comes from its deterministic rules. Across these designs the model reads the heterogeneous
fields and organizes the evidence; the pairing itself is still decided by logs,
amounts, times, and rules.

\smallskip
\emph{\underline{Insight 2}: In every surveyed work, the role of LLMs and agents is to lower
adaptation cost; the final decision is still made by deterministic rules over
verifiable evidence.}

\subsection{Cross-Mechanism Analysis: The Evidence Hierarchy}
\label{sec:matching-insights}

Ranked by evidence strength, the three mechanisms form a chain that degrades step
by step: an identifier gives the pair directly, generic fields only shrink the
candidate set, and models serve the remaining bridges whose field semantics vary.
Systems therefore use strong evidence first and fall back only where it is
missing: Jigsaw~\cite{hu2024jigsaw}
pairs with the identifier on bridges whose events carry a unique field and uses
heuristics on the rest, and the arbitrage study~\cite{oz2025arbitrage} joins on the
message number on native bridges such as Arbitrum, Base, and Optimism, falling back
to token-transfer scanning on Polygon, which leaves none
(the Mech column of Table~\ref{tab:matching} records these combinations). The gap
in matching rates is also set by the evidence level. Within one heuristic
framework, for example, the Ethereum-Polygon tracing~\cite{yan2025tracing}
reaches 92.78\% on ERC721 withdrawals but 67.55\% on ERC20 withdrawals
(\S\ref{sec:heuristics}).

\smallskip
\emph{\underline{Insight 3}: The three matching mechanisms are ordered by the strength of their
evidence; the ceiling of matching performance is set by the evidence the
cross-chain system leaves behind.}

\subsection{Downstream Security Applications}

Matching results feed two security tasks: attack detection and fund
tracing. Jigsaw's CrossAlert~\cite{hu2024jigsaw} runs three rules over the
matching results of about 80.05 million transactions and flags 94 abnormal
matching transactions, of which 47 unmatched ones involve over \$605M in losses;
ABCTRACER~\cite{lin2025abctracer} identifies 20 attack transaction pairs and
separately recovers 10 laundering-related cross-chain paths;
ConneX~\cite{liang2025connex} traces Bybit-related stolen funds across a bridge
to Solana. These applications also leave three issues unsettled. First, detectors disagree on
whether pairing should be a premise: BridgeGuard~\cite{wu2025bridgeguard} does
not require pairing, finding that 65.7\% of attack transactions have no
counterpart on the other chain, whereas BridgeShield~\cite{lin2025bridgeshield}
builds its detection graph on paired transactions. Second, a missing counterpart
is ambiguous: it can be the mark of an attack (a fake deposit releases assets
with no real deposit behind them~\cite{lin2024fakedeposit}) or a gap in the
collected data (\S\ref{sec:challenges}). Third, evaluation is not
comparable across these systems: label sources and sample units differ, and
systems that claim real-time operation report throughput but no end-to-end alert
latency~\cite{wiputra2025bads,tran2025veribridge}.

\smallskip
\emph{\underline{Insight 4}: Matched pairs feed attack detection and fund tracing, but
whether pairing should be a detection premise and how a missing counterpart
should be read, remain disputed.}

\section{Dataset Availability and Reproducibility}
\label{sec:datasets}

This section compares the cross-chain transaction datasets built by existing research
along coverage, scale, label source, and actual availability. We checked each paper's
stated data or artifact release one by one, including dataset links, code
repositories, and external data sources, and rated availability on the two levels
defined in the notes of Table~\ref{tab:datasets} (checked July 2026).

\providecommand{\grayrule}{\arrayrulecolor{black!15}\specialrule{0.3pt}{1pt}{1pt}\arrayrulecolor{black}}

\begin{table*}[t]
\centering
\begin{threeparttable}
\caption{Cross-Chain Transaction Datasets, Sorted by Verified Availability (as of July 2026)}
\label{tab:datasets}
\scriptsize
\setlength{\tabcolsep}{4pt}
\renewcommand{\arraystretch}{1.1}

\begin{tabular}{@{}r p{2.5cm} p{2cm} p{2.5cm} p{2.5cm} p{2.5cm} @{}}
\toprule
\multicolumn{1}{c}{Work}
& \multicolumn{1}{c}{Scope}
& \multicolumn{1}{c}{Size}
& \multicolumn{1}{c}{Period}
& \multicolumn{1}{c}{Label source}
& \multicolumn{1}{c}{Availability} \\
\midrule
\quad\quad XChainDataGen~\cite{augusto2025xchaindatagen} \quad\quad
& 5 bridges, 11 chains
& 11.29M pairs
& 2024.06--2024.12
& on-chain identifier
& \cmark GitHub+Zenodo \\
\grayrule
 \quad\quad Empirical Study~\cite{yan2025empirical}  \quad\quad
& 4 bridges, 11 chains
& 543K pairs
& 2023
& platform record
& \cmark Zenodo \\
\grayrule
 \quad\quad XChainWatcher~\cite{augusto2025xchainwatcher}  \quad\quad
& 2 bridges, 3 chains
& 81K pairs
& 2022
& attack event
& \cmark GitHub \\
\grayrule
 \quad\quad CONNECTOR~\cite{lin2025connector}  \quad\quad
& 3 bridges, 3 chains
& 24K pairs
& 2020.09--2023.04
& platform record
& \cmark GitHub \\
\grayrule
 \quad\quad ABCTRACER~\cite{lin2025abctracer}  \quad\quad
& 12 bridges, 2 chains
& 29K pairs
& 2021.04--2024.03
& platform record
& \cmark GitHub \\
\grayrule
 \quad\quad LOCARD~\cite{yu2026locard}  \quad\quad
& 1 bridge, 4 chains
& 151K pairs
& 2025
& platform record
& \cmark GitHub \\
\grayrule
 \quad\quad Price of Interop.~\cite{cao2026priceinteroperability}  \quad\quad
& 16 bridges, 20 chains
& 9M pairs
& 2022--2025
& --
& \cmark GitHub \\
\midrule
 \quad\quad Jigsaw~\cite{hu2024jigsaw}  \quad\quad
& 13 bridges, 7 chains
& 80.05M tx
& 2021.03--2024.01
& --
& \xmark promised only \\
\grayrule
 \quad\quad ICE~\cite{hu2024instant}  \quad\quad
& 9 bridges
& --
& 2023.01--2023.09
& --
& \xmark no link \\
\grayrule
 \quad\quad ConneX~\cite{liang2025connex}  \quad\quad
& 5 bridges
& 504K tx
& 2021.02--2024.03
& platform record
& \xmark expired anon repo \\
\bottomrule
\end{tabular}

\begin{tablenotes}[flushleft]
\scriptsize
\item[-] Size: ``pairs'' denotes linked deposit--withdrawal units, while ``tx'' denotes individual one-sided transactions.
\item[-] For availability, \cmark indicates that the data and key labels remain obtainable, whereas \xmark indicates that the dataset was claimed to be public but its link is missing or empty, or points to an expired anonymous repository.
\item[-] Works that make no public-release claim are discussed in the text.
\end{tablenotes}
\end{threeparttable}
\end{table*}

Seven of the fifteen surveyed works provide an obtainable data artifact, fewer
than half, and not every artifact carries task labels. The failures fall into
three modes. The first is promised-only: Jigsaw~\cite{hu2024jigsaw} states that
it will release its eighty-million-transaction dataset, but no link can be found
now. The second is a missing link: the ICE study~\cite{hu2024instant} calls its dataset
open-source but gives no URL anywhere in the paper. The third is an expired
anonymous review repository: ConneX~\cite{liang2025connex} placed its data in an
anonymous repository built for double-blind review, and once it expired the data
was gone. The remaining five papers make no public-release claim. Among them,
XSema~\cite{zheng2024xsema} states that it built ``the first cross-chain semantic
dataset'' yet gives no way to obtain it.

Among the truly obtainable datasets, the main difference lies in the label source.
The matching ground truth of Yan et
al.~\cite{yan2025empirical}, CONNECTOR~\cite{lin2025connector},
ABCTRACER~\cite{lin2025abctracer}, and LOCARD~\cite{yu2026locard} comes from bridge
ledgers, explorers, or official execution records, which ultimately depend on
platform cooperation;
XChainDataGen~\cite{augusto2025xchaindatagen} builds its cctx ground truth by
joining deposit and withdrawal events
on the identifiers the bridge contracts write on chain, so it does not depend on
platform cooperation. Beyond matching truth,
XChainWatcher~\cite{augusto2025xchainwatcher} marks the transactions involved in
two real bridge attacks (Ronin and Nomad), a label that serves downstream
detection, and the
Price of Interoperability dataset~\cite{cao2026priceinteroperability} carries no
security labels at all. In coverage, public datasets concentrate
on the EVM ecosystem and the 2021--2025 period: LOCARD's dataset is the only one targeting
native-asset chains (BTC, DOGE, LTC), and the ICE and cross-ledger-service direction
has no usable public dataset. Overall, a cross-chain dataset stays reproducible only
as long as every source it depends on stays available: the archive that stores it,
the platform records behind its ground truth, and the chain data itself. Future works should archive all three together, so that a matching
result can still be re-checked after the platform changes or disappears.

\section{Open Challenges}
\label{sec:challenges}

Building on the systematization above, we summarize four open challenges.

\textit{Interpreting missing counterparts under partial evidence.}
When a transaction's counterpart cannot be found, there are two possible reasons:
the counterpart exists but lies outside the collected data, or it never existed.
The second is exactly what a fake deposit looks like: the destination chain
releases an asset although no real deposit was ever made
(\S\ref{sec:matching}). Future matching should make the distinction
explicit: check how well the collected data covers the bridge, the chains, and
the time window in question, and report
how likely the missing counterpart is a data gap rather than an attack, instead of
a hard verdict. How much a missing counterpart means also depends on how complete
the matching is. When identifiers match nearly every transaction, as in
Jigsaw~\cite{hu2024jigsaw}, the few that remain unmatched deserve individual
inspection; when heuristics leave a third of withdrawals unmatched
(\S\ref{sec:heuristics}), an unmatched withdrawal by itself signals
nothing. A verdict on a missing counterpart should therefore state how complete
the matching behind it is.

\textit{Matching one-to-many and many-to-many transfers.}
Existing matching generally assumes one deposit corresponds to one withdrawal
(\S\ref{sec:matching}), while aggregators~\cite{subramanian2024aggregators},
liquidity pools, and pooled settlement produce
one-to-many and many-to-many correspondences that weaken amount conservation, the
time window, and address constraints at the same time. Correspondences that span
several pairs also occur: scanning 34.8 million bridge transactions across twelve
networks, the N-hop study~\cite{mancino2025bunnyhops} finds ten arbitrage routes
that cross two or three bridges in a row, and such a route is recovered only when
the time and amount constraints hold at every hop. Future matching should relax
one-to-one matching into set-to-set matching: constraints such as amount
conservation should be checked over sets of
transactions, candidate combinations scored by evidence strength, and multiple
candidates reported with confidence instead of a fixed rule selecting one and
discarding the alternatives.

\textit{Ground truth, benchmarks, and artifact availability.}
Matching ground truth ranges from official platform records to attack events, and
a substantial share of artifacts claimed to be public are unobtainable
(\S\ref{sec:datasets}). The most direct fix is durable storage: data and
labels should live in archives that outlast the platform and the review process
(the Zenodo archives of Yan et al.~\cite{yan2025empirical} and
XChainDataGen~\cite{augusto2025xchaindatagen} are ready examples). Two papers
that both report a matching rate are not comparable if their ground-truth sources
differ. A shared test set would fix this. It should hold transactions from
several bridges; each matched pair should state where its label came from (a
platform record, an on-chain identifier, or an attack report); and a few bridges
should be held back from development, so every method also reports how it does on
a bridge it has never seen. XSema~\cite{zheng2024xsema} and
ABCTRACER~\cite{lin2025abctracer} already test on unseen bridges, but each on its data, so their numbers still cannot be compared. Until a shared set exists, releasing the ground truth next to the code, as
the seven obtainable artifacts in Table~III do, is the practical minimum.

\textit{Verifiable use of models and agents.}
Existing systems confine LLMs and agents to components: field discovery and
forensic orchestration. The final conclusion still rests on
deterministic rules, economic constraints, or auditable evidence
(\S\ref{sec:matching}). The open question is whether such a component adds
enough value over a deterministic script or a human-written procedure to outweigh
its nondeterminism and the extra verification it requires. Future work should
evaluate the component on three dimensions: task quality (does precision or F1
improve), adaptation cost (do rules, annotation, and analysis time decrease), and
auditability (can the prompts, tool calls, and final judgment be replayed and
independently checked). No surveyed system measures all three.

\section{Conclusion}
\label{sec:conclusion}

This paper systematizes research on cross-chain transaction identification and
matching. We organize identification into four approaches and matching into
three mechanisms whose performance is bounded mainly by the strength of the
evidence the underlying system leaves, summarize what matched pairs supply to
downstream attack detection and fund tracing, verify the actual availability of
existing datasets and artifacts one by one, and distill four open challenges.
The core bottleneck of cross-chain transaction analysis lies beyond the choice
of algorithm: whether the evidence exists, whether the ground truth is credible,
whether the data is reproducible, and whether analysis can deliver a verifiable
judgment under incomplete evidence. Future work should move from offline case
studies toward evaluation that answers these four questions directly.

\bibliographystyle{IEEEtran}
\bibliography{ref}

\end{document}